\documentclass{elsart}

\usepackage{natbib}

\usepackage{amssymb}
\usepackage{graphicx}

\begin{document}

\begin{frontmatter}



\title{Evolution of the periodicities in 2S 0114+650}

\author{Ravi Sood$^a$, Sean Farrell$^a$, Paul O'Neill$^b$, Ravi Manchanda$^c$, N. M. Ashok$^d$}

\address{$^a$School of PEMS, UNSW@ADFA, Canberra 2600, Australia\\
$^b$Astrophysics Group, Imperial College London, London SW7 2AZ,
UK\\
$^c$Tata Institute of Fundamental Research, Mumbai 400005, India\\
$^d$Physical Research Laboratory, Navrangpura, Ahmedabad 380009,
India}

\begin{abstract}
% Text of abstract
We have analysed nine years of data from the All Sky Monitor on the
\textit{Rossi X-ray Timing Explorer} for
2S 0114+650 to study the evolution of its
spin, binary and super-orbital periods.  The spin history of the
neutron star in this system exhibits torque reversals lasting $\sim$
1 yr.  The newly discovered super-orbital period has remained stable
over the 9-yr span, making 2S 0114+650 the fourth known system to
exhibit stable super-orbital modulation. We compare its super-orbital
period evolution with those of the other three such systems.

\end{abstract}

\begin{keyword}
X-rays: binary  -  stars:  neutron  -  accretion:  accretion discs -
stars: individual (2S 0114+650)

\end{keyword}

\end{frontmatter}

% main text
\section{Introduction}
The high mass X-ray binary  2S 0114+650 (Per X-2) was discovered in
1977,
 and subesquently the optical donor star LSI +65$^{\circ}$ 010 was classified
as a B1 Ia supergiant at a distance of 7.0 $\pm$ 3.6 kpc (Reig et
al. 1996). Ashok et al. (2006) recently confirmed this
classification through near infrared observations at the Mt Abu
observatory. The binary orbital period of 11.6 d was determined in
X-rays by Corbet, Finley $\&$ Peele (1999) using data from the All
Sky Monitor (ASM) aboard the \textit{Rossi X-ray Timing Explorer
(RXTE)}. They also confirmed a periodicity of $\sim$2.7 h,
attributing it to emission from a highly magnetised neutron star.
The 2.7 h spin period is by far the slowest known for an X-ray
pulsar, making 2S 0114+650 the first in a new class of super-slow
rotators. Farrell, Sood $\&$ O'Neill (2006) recently reported a 30.7
d super-orbital period in $\sim$8.5 yr of $\textit{RXTE}$ ASM data,
making 2S 0114+650 the fourth X-ray binary exhibiting a stable
super-orbital modulation.

\section{Data Reduction $\&$ Analysis}
Archived data from the \textit{RXTE} ASM for the X-ray binaries 2S
0114+650, SS433, Her X-1, and LMC X-4 for the period MJD 50135 (1996
February 22) to MJD 53635 (2005 September 22), were used in these
analyses. We used one-day average light curves for the analysis of
the orbital and super-orbital periods, and 90 s dwell data to
investigate the evolution of the spin period (Farrell et al. 2006).
A Dynamic Power Spectrum approach was used for the orbital and
super-orbital periods (e.g. Clarkson et al. 2003) where the
Lomb-Scargle periodogram was applied to overlapping light curve
sections, producing a sliding `data window' sensitive to long-term
changes in any modulation present in the light curves. We used a
window length of 400 d and a shift of 100 d for the analysis of the
orbital and super-orbital periods. For the 164 d super-orbital
period of SS433 we used a window length of 800 d. C-band data (5.0 -
12.0 keV range) were used for 2S 0114+650 and SS433 to account for
the excessive noise observed in their low energy channels. The
combined 1.5 - 12.0 keV data were used for the other sources. The
resulting power spectra were normalised to the highest power in the
period range of interest.

\begin{figure}[h!]
\begin{center}
\includegraphics[width=9cm]{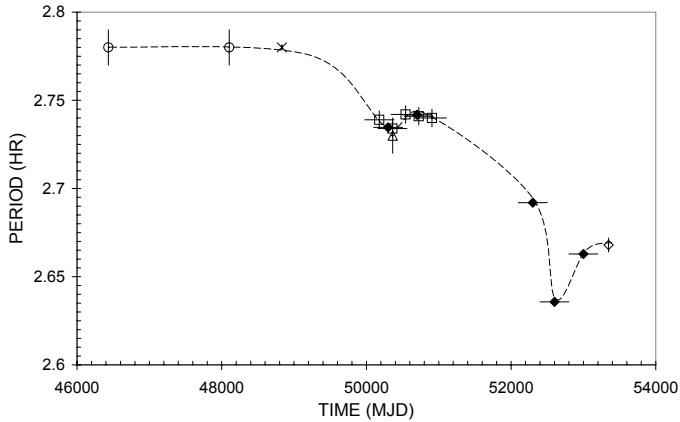}
\caption{The evolution of the $\sim$2.7 h spin period of 2S
0114+650. See Farrell et al. (2006) for details of where the
individual data points were taken from.}
\end{center}
\end{figure}

\section{Results $\&$ Discussion}
The neutron star spin evolution in 2S 0114+650 is shown in Fig. 1. Data
points are shown for only when the spin period power exceeded the
99$\%$ significance level for this low flux source.
Two episodes of torque reversal, each lasting $\sim$ 400 d, are
embedded in a general spin-up trend.  2S 0114+650 is expected to be
a wind-fed accretor, and the association of the torque reversal
episodes with the formation of a transient accretion disc is
unclear. The observed average $\dot{P}$/$P$ based on \textit{RXTE}
ASM data is 3.4 $\times$ 10$^{-3}$ yr$^{-1}$, a value which has been
independently confirmed by Bonning $\&$ Falanga (2005) using \textit{INTEGRAL}
observations. This value is an order of
magnitude less than the value expected when the spin
up is caused by a disc-fed system (Nagase 1989).

\begin{figure}[h]
\begin{center}
\includegraphics[width=10.9cm]{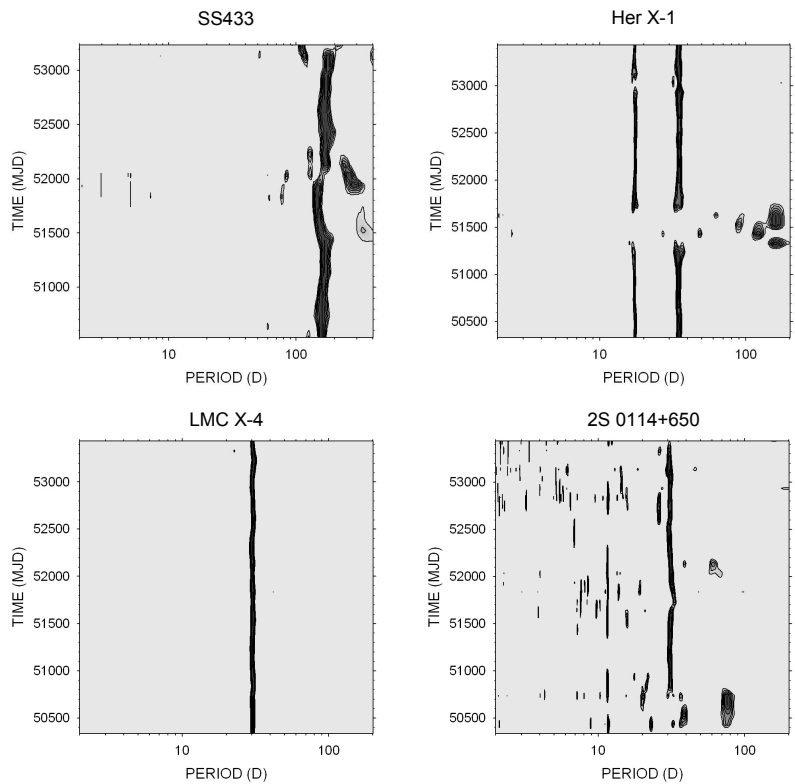}
\caption{Dynamic Power Spectra of the four stable super-orbital
periods.}
\end{center}
\end{figure}

\begin{figure}[h]
\begin{center}
\includegraphics[width=11cm]{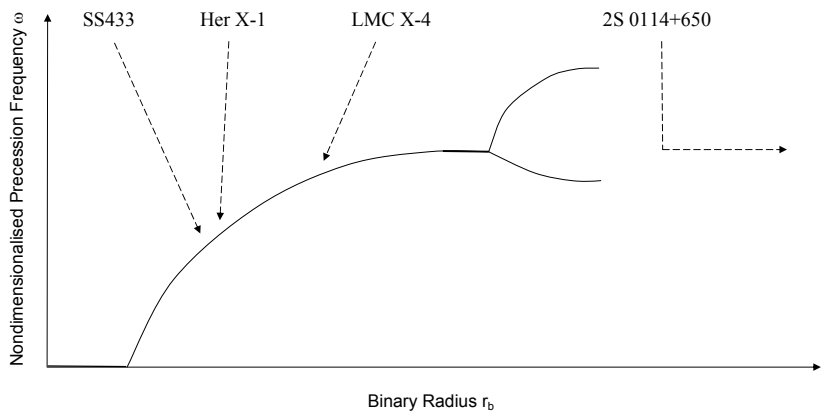}
\caption{Schematic bifurcation diagram for radiation-driven warping
(Figure 14 from Clarkson et al. (2003)).
With a binary radius of 18.8 $\times$
10$^{6}$ GM$_{1}$c$^{-2}$, 2S 0114+650 lies far off to the right of
this diagram.}
\end{center}
\end{figure}

The change in the orbital period cannot be tightly
constrained by the ASM data for 2S 0114+650.  An upper limit value for
$\dot{P}_{orb}$/$P_{orb}$ of $\lesssim$10$^{-3}$ yr$^{-1}$ is obtained.

To date four stable super-orbital periods have been identified in
X-ray binaries: SS433, Her X-1, LMC X-4, and now 2S 0114+650. The 164 d period in
SS433 has been explained by jet precession (Margon 1984).
Clarkson et al. (2003)
showed that the stability of the
35 d and 30 d super-orbital periods in Her X-1 and LMC X-4
were consistent with the radiation-driven
warping of an accretion disc. They demonstrated in Fig. 14 of their paper
that the number of stable precession
solutions was dependent on the binary separation. SS433, Her X-1 and
LMC X-4 all lie in the stable mode-0 precession region, while 2S
0114+650 lies far
to the right of the mode 1 region (Fig. 3), indicating that two or more
steady solutions are possible and that the system should precess at
a combination of these warping modes. Thus, if the 30.7 d
super-orbital period in 2S 0114+650 is due to the precession of a
radiatively warped accretion disc, it should not exhibit the
long-term stability that has been observed.

\section{Acknowlegements}
This research made use of quick-look data provided by the
$\textit{RXTE}$ ASM team at MIT and GSFC and data obtained through
the High Energy Astrophysics Science Archive Research Centre Online
Service, provided by NASA/GSFC.

% \appendix

% \section{}
% \label{}

% Bibliographic references with the natbib package:
% Parenthetical: \citep{Bai92} produces (Bailyn 1992).
% Textual: \citet{Bai95} produces Bailyn et al. (1995).
% An affix and part of a reference:
%   \citep[e.g.][Ch. 2]{Bar76}
%   produces (e.g. Barnes et al. 1976, Ch. 2).

\end{document}